\documentclass[11pt]{article}
\usepackage{color}
\usepackage{url}
\usepackage{hyperref}
\usepackage{amsmath}
\usepackage{amsfonts}
\usepackage{amssymb}
\usepackage{amsthm}
\usepackage{geometry}
\usepackage{enumerate}
\ifdefined\directlua
\usepackage{polyglossia}
\setmainlanguage{english}

\usepackage{fontspec}
\usepackage{microtype}
\usepackage{float}
\restylefloat{table}
\usepackage{pdftexcmds}
\makeatletter
\ifcase\pdf@shellescape
\relax\or
\usepackage{luacode}
\begin{luacode}
  function prtgit()
  local cmd="git show -s --format='
  print(cmd)
  local r=io.popen(cmd):read("*a")
  if (r) then
  tex.print([[\string\def\string\COMMIT{]]..r..[[}]])
  end
  end
\end{luacode}
\directlua{prtgit()}
\or
\relax\fi
\makeatother
\ifdefined\COMMIT
        \usepackage{background}
        \backgroundsetup{%
         pages=all, placement=bottom,angle=0,scale=1.6,%
         vshift=20pt,hshift=0pt,
         contents={Commit version:\COMMIT}}
\fi
\fi

\newcommand{\cL}{\mathcal L}
\newcommand{\cG}{\mathcal G}

\newcommand{\F}{\mathbb F}

\newcommand{\C}{\mathcal C}
\newcommand{\D}{\mathcal D}

\newcommand{\cH}{\mathcal H}

\newcommand{\N}{\mathrm{N}}

\newcommand{\Hom}{\mathrm{Hom}}

\newtheorem{theorem}{Theorem}[section]
\newtheorem{lemma}[theorem]{Lemma}

\newtheorem{proposition}[theorem]{Proposition}

\theoremstyle{definition}

\title{$\mathbb{F}_{q^n}$-linear rank distance codes and their distinguishers\thanks{{The
research  was supported by
Ministry for Education, University and Research of Italy MIUR (Project
PRIN 2012 "Geometrie di Galois e strutture di incidenza") and by the Italian National
Group for Algebraic and Geometric Structures and their Applications (GNSAGA
- INdAM).}}}
\author{Luca Giuzzi and Ferdinando Zullo}
\date{}
\begin{document}
\maketitle

\begin{abstract}
  For any admissible value of the parameters there exist
  Maximum Rank distance (shortly MRD) $\F_{q^n}$-linear codes of $\F_q^{n\times n}$ (see Subsection \ref{Rep} for the definition).
  It has been shown in \cite{H-TNRR} (see also \cite{ByrneRavagnani}) that, if field extensions large enough are
  considered, then \emph{almost all} (rectangular) rank distance codes are MRD.
  On the other hand, very few families of $\F_{q^n}$-linear codes
  are currently known up to equivalence.
  One of the possible applications of MRD-codes is for McEliece--like public
  key cryptosystems, as proposed by Gabidulin, Paramonov and Tretjakov in \cite{GPT}. In this framework it is very important to obtain
  new families of MRD-codes endowed with fast decoding algorithms.
  Several decoding algorithms exist for Gabidulin codes as shown in \cite{Gabidulin}, see also \cite{Loi06,PWZ,WT}.
  In this work, we will survey the known families of $\F_{q^n}$-linear MRD-codes,   study some invariants of MRD-codes and evaluate their value for
  the known families, providing  a  characterization of
  generalized twisted Gabidulin codes as done in \cite{GiuZ}.

\smallskip

\end{abstract}

\section{Introduction}

Delsarte \cite{Delsarte} introduced   rank-distance (RD) codes in 1978 as the
$q$-analogs of the usual linear error correcting codes endowed with Hamming distance.
The set of $m \times n$ matrices $\F_q^{m\times n}$ over $\F_q$ is a rank metric $\F_q$-space where the rank metric distance is given by
\[d(A,B) = \mathrm{rk}\,(A-B)\]
for $A,B \in \F_q^{m\times n}$.

A subset $\C \subseteq \F_q^{m\times n}$ is called a \emph{rank distance code} (RD-code for short). The minimum distance of $\C$ is
\[d(\C) = \min_{{A,B \in \C},\ {A\neq B}} \{ d(A,B) \}.\]
We say that $\C$ has parameters $(n,m,q;d)$, where $d$ is the minimum distance of $\C$.
When $\C$ is an $\F_q$-linear subspace of $\F_q^{m\times n}$, we say that $\C$ is an $\F_q$-\emph{linear}\index{linear RD-code} RD-code and its dimension $\dim_{\F_q}\C$ is defined the dimension of $\C$  as a subspace over $\F_q$.
In \cite{Delsarte}, Delsarte also showed that the
parameters of these RD-codes must obey a
Singleton-like bound:

\begin{theorem}
Let $\C$ be an RD-code of $\F_q^{m\times n}$ and let $d$ its minimum distance, then
\[ |\C| \leq q^{\max\{m,n\}(\min\{m,n\}-d+1)}. \]
\end{theorem}

When the equality holds, $\C$ is a \emph{maximum rank distance} (\emph{MRD}\index{MRD-code} for short) code.
Examples of MRD-codes were first found by Delsarte in \cite{Delsarte} and rediscovered by Gabidulin in \cite{Gabidulin} and by Roth in \cite{Roth91}.
For $m\neq n$ two RD-codes $\C$ and $\C'$ are said to be equivalent if and only if there exist $X \in \mathrm{GL}(m,q)$, $Y \in \mathrm{GL}(n,q)$, $Z \in \F_q^{m\times n}$ and a field automorphism $\sigma$ of $\F_q$ such that
\[\C'=\{XC^\sigma Y + Z \colon C \in \C\}.\]
When $m=n$ we have two possible definitions:
\begin{enumerate}
  \item $\C$ and $\C'$ are equivalent if there are invertible matrices $X,Y \in \F_q^{n\times n}$, $Z \in \F_q^{n \times n}$ and a field automorphism $\sigma$ of $\F_q$ such that
      \[ \C'=\{XC^\sigma Y + Z \colon C \in \C\}. \]
  \item $\C$ and $\C'$ are equivalent if there are invertible matrices $A,B \in \F_q^{n\times n}$, $Z \in \F_q^{n \times n}$ and a field automorphism $\sigma$ of $\F_q$ such that
      \[\C'=\{XC^\sigma Y + Z \colon C \in \C\} \,\, \text{or} \,\, \C'=\{X(C^t)^\sigma Y + Z \colon C \in \C\}. \]
\end{enumerate}
From now on when we will talk about the equivalence between two RD-codes of $\F_q^{n\times n}$ we will mean the first kind of equivalence mentioned above, otherwise we will talk about \emph{strong equivalence}.
Clearly, when $\C$ and $\C'$ are $\F_q$-linear, we may always assume that $Z$ is the zero matrix.
For further details on the equivalence of RD-codes see also \cite{Berger,Morrison}.

\medskip

Let $\C\subseteq \F_{q}^{m \times n}$ be an RD-code; the \emph{adjoint code}\index{adjoint code} of $\C$ is
\[ \C^\top =\{C^t \colon C \in \C\}. \]

Consider the symmetric bilinear form $\langle\cdot,\cdot\rangle$ on $\F_q^{m \times n}$ given by
\[ \langle M,N \rangle= \mathrm{Tr}(MN^t). \]
The \emph{Delsarte-dual code} of an $\F_q$-linear RD-code $\C$ is
\[ \C^\perp = \{ N \in \F_q^{m\times n} \colon \langle M,N \rangle=0 \, \text{for each} \, M \in \C \}. \]

By using the machinery of association schemes, Delsarte in \cite{Delsarte} proved the following result.

\begin{lemma}\label{dualMRD}\cite[Theorem 5.5]{Delsarte}
Let $\C\subseteq \F_{q}^{m \times n}$ be an $\F_q$-linear MRD-code of dimension $k$ with $d>1$. Then the Delsarte dual code $\C^\perp\subseteq \F_{q}^{m\times n}$ is an MRD-code of dimension $mn-k$.
\end{lemma}

An elementary proof of the above result can be found in \cite{Ravagnani}.

\medskip

In general, it is difficult to determine whether two RD-codes are equivalent or not.
The notion of \emph{idealiser} provides a useful criterion.

\smallskip

Let $\C\subset \F_q^{m\times n}$ be an RD-code; its \emph{left} and \emph{right idealisers}
$L(\C)$ and $R(\C)$ are defined as
\[ L(\C)=\{ Y \in \F_q^{m \times m} \colon YC\in \C\hspace{0.1cm} \text{for all}\hspace{0.1cm} C \in \C\}\]
\[ R(\C)=\{ Z \in \F_q^{n \times n} \colon CZ\in \C\hspace{0.1cm} \text{for all}\hspace{0.1cm} C \in \C\}.\]
These notions have been introduced by  Liebhold and Nebe in \cite[Definition 3.1]{LN2016}.
The idealisers are subgroups of the automorphism group, which are easier to calculate.
Such sets appear also in the paper of Lunardon, Trombetti and Zhou \cite{LTZ2}, where they are
respectively called \emph{middle nucleus} and \emph{right nucleus};
therein the authors investigate these sets and, in particular,
prove the following results.

\begin{proposition}\cite[Proposition 4.1]{LTZ2}\label{idealis}
  If $\C_1$ and $\C_2$ are equivalent $\F_q$-linear RD-codes of $\F_q^{m\times n}$, then their
  left (resp. right) idealisers are also equivalent as RD-codes.
\end{proposition}

\begin{theorem}\cite[Theorem 5.4 \& Corollary 5.6]{LTZ2}\label{finite}
Let $\C$ be an $\F_q$-linear MRD-code of $\F_q^{m\times n}$ with minimum distance $d>1$.
If $m \leq n$, then $L(\C)$ is a finite field with $|L(\C)|\leq q^m$.
If $m \geq n$, then $R(\C)$ is a finite field with $|R(\C)|\leq q^n$.
In particular, when $m=n$ then $L(\C)$ and $R(\C)$ are both finite fields.
\end{theorem}

\subsection{Representation of RD-codes as $q$-polynomials}\label{Rep}

Any $\F_q$-linear RD-code over $\F_q$ can
be equivalently regarded either as a subspace of matrices in $\F_{q}^{m\times n}$
or as a subspace of $\Hom(V_n,V_m)$. In the present section we
shall recall a specialized representation for these codes in terms of
linearized polynomials which we shall
use in the rest of this abstract.

Consider two vector spaces $V_n$ and $V_m$ over $\F_q$ with dimension respectively $n$ and $m$. If $n\geq m$
we can always regard $V_m$ as a subspace of $V_n$ and identify
$\Hom(V_n,V_m)$ with the subspace of those $\varphi\in\Hom(V_n,V_n)$ such that
$\mathrm{Im}(\varphi)\subseteq V_m$. Also, $V_n\cong\F_{q^n}$,
when $\F_{q^n}$ is considered as an $\F_q$-vector space of dimension $n$.
Let now  $\Hom_q(\F_{q^n}):=\Hom_q(\F_{q^n},\F_{q^n})$ be the set of all
$\F_q$--linear maps of $\F_{q^n}$ in itself.
It is well known that each element of $\Hom_q(\F_{q^n})$ can be
represented in a unique way as a $q$--polynomial over $\F_{q^n}$;
see e.g. \cite{lidl_finite_1997}.
In other words, for any $\varphi\in\Hom_q(\F_{q^n})$ there is
an unique polynomial $f(x)$ of the form
\[ f(x):=\sum_{i=0}^{n-1} a_i x^{q^i}\]
with $a_i \in \F_{q^n}$ such that
\[ \forall x\in\F_{q^n}:\varphi(x)=f(x)=a_0x+a_1x^q+\cdots+a_{n-1}x^{q^{n-1}}. \]
The set $\tilde{\cL}_{n,q}$ of such $q$-polynomials over $\F_{q^n}$ is
a vector space over $\F_{q^n}$ of dimension $n$ with respect to the
usual sum and scalar multiplication.
When it is regarded as a
vector space over $\F_q$, its dimension is $n^2$
and it is isomorphic to $\F_q^{n\times n}$.
Actually, $\tilde{\cL}_{n,q}$ endowed with the product $\circ$
induced by the functional
composition in $\Hom_q(\F_{q^n})$ is an algebra over $\F_{q}$.

Using the above remarks,
it is straightforward to see that any
$\F_q$-linear RD-code might be regarded as a suitable $\F_q$-subspace
of $\tilde{\cL}_{n,q}$, when $m=n$. This approach shall be extensively used  in the present
work. In order to fix the notation and ease the reader, we shall
reformulate some of the notions recalled before in terms of $q$-polynomials.
Also, in the remainder of this work we shall
always silently identify the elements of $\tilde{\cL}_{n,q}$ with the
morphisms of $\Hom_q(\F_{q^n})$ they represent and
speak also of \emph{kernel} and \emph{rank} of a linearized polynomial.

The notion of Delsarte dual code can be written in terms of
$q$-polynomials as follows,
see for example \cite[Section 2]{LTZ}.
Let $b:\tilde{\cL}_{n,q}\times\tilde{\cL}_{n,q}\to\F_q$ be the bilinear form
given by
\[ b(f,g)=\mathrm{Tr}_{q^n/q}\left( \sum_{i=0}^{n-1} f_ig_i \right) \]
where $\displaystyle f(x)=\sum_{i=0}^{n-1} f_i x^{q^i}$ and $\displaystyle g(x)=\sum_{i=0}^{n-1} g_i x^{q^i} \in \F_{q^n}[x]$ and
denote by $\mathrm{Tr}_{q^n/q}$  the trace function $\F_{q^n}\to\F_q$.
The Delsarte dual code $\C^\perp$ of a set of $q$-polynomials $\C$ is
\[\C^\perp = \{f \in \tilde{\cL}_{n,q} \colon b(f,g)=0, \hspace{0.1cm}\forall g \in \C\}. \]
The adjoint code $\C^\top$ of a set of $q$-polynomials $\C$ is
\[ \C^\top= \{\hat{f} \in \tilde{\cL}_{n,q} \colon f \in \C\}, \]
where $\hat{f}$ is the adjoint of $f(x)$ with respect to the bilinear form $b$.

Also, two RD-codes $\C$ and $\C'$ are equivalent if and only if
there exist two invertible $q$-polynomials  $h$ and $g$ and a field automorphism $\sigma$ such that
$\{h \circ f^\sigma \circ g \colon f\in \C\}=\C'$, where if $\displaystyle f(x)=\sum_{i=0}^{n-1}a_ix^{q^i}$ then $\displaystyle f^\sigma(x)=\sum_{i=0}^{n-1}a_i^\sigma x^{q^i}$; furthermore, $\C$ and $\C'$ are strongly equivalent if and only if $\C'$ is equivalent to either $\C$ or to $\C^\top$.

The left and right idealisers of a code
$\C\subseteq\tilde{\cL}_{n,q}$ can be written as
\[L(\C)=\{\varphi(x) \in \tilde{\cL}_{n,q} \colon \varphi \circ f \in \C\, \text{for all} \, f \in \C\};\]
\[R(\C)=\{\varphi(x) \in \tilde{\cL}_{n,q} \colon f \circ \varphi \in \C\, \text{for all} \, f \in \C\}.\]

Let $\C$ be an $\F_q$-linear MRD-code of dimension $nk$ with parameters {$(n,n,q;n-k+1)$}.
If $L(\C)$ has maximum cardinality $q^n$, then we may always assume (up to equivalence) that
\[L(\C)=\mathcal{F}_n=\{\tau_{\alpha}=\alpha x \colon \alpha \in \F_{q^n}\}\simeq \F_{q^n};\]
the same holds for the right idealiser (see e.g. \cite{Huppert}).

Hence, when the left idealiser is $\mathcal{F}_n$, $\C$ turns out to be closed with respect to the left composition with $\F_{q^n}$-linear maps, while if the right idealiser is $\mathcal{F}_n$, then $\C$ is closed with respect to the right composition with the $\F_{q^n}$-linear maps.
For this reason, when $L(\C)$ (resp. $R(\C)$) is equal to $\mathcal{F}_n$ we say that $\C$ is $\F_{q^n}$-\emph{linear on the left} (resp. \emph{right}) (or simply $\F_{q^n}$-\emph{linear} if it is clear from the context).
In the literature it is quite common to find the term $\F_{q^n}$-linear instead of $\F_{q^n}$-linear on the left.
We can state the following result.

\begin{proposition}\cite[Theorem 6.1]{CMPZ}\cite[Theorem 2.2]{CsMPZh}\label{rightvectorspace}
Let $\C$ be an $\F_q$-linear MRD-code of dimension $nk$ with parameters $(n,n,q;n-k+1)$.
Then $L(\C)$ (resp. $R(\C)$) has maximum order $q^n$ if and only if there exists an MRD-code $\C'$ equivalent to $\C$ which is $\F_{q^n}$-linear on the left (resp. on the right).
\end{proposition}

For more details see also \cite{Zullo}.

\section{Known examples of $\F_{q^n}$-linear MRD-codes}\label{examples}

This section is devoted to list the known examples of MRD-codes, by using their representation as sets of linearized polynomials of $\tilde{\mathcal{L}}_{n,q}$.
In \cite{Delsarte}, Delsarte gives the first construction for linear MRD-codes (he calls such sets \emph{Singleton systems}) from the perspective of bilinear forms. Few years later, Gabidulin in \cite[Section 4]{Gabidulin} presents the same class of MRD-codes by using linearized polynomials.
Although these codes have been originally discovered by Delsarte, they are called \emph{Gabidulin codes} and they can be written as follows
\[ \mathcal{G}_{k}=\langle x,x^q,\ldots,x^{q^{k-1}} \rangle_{\F_{q^n}}, \]
with $k\leq n-1$.
Kshevetskiy and Gabidulin in \cite{kshevetskiy_new_2005} generalize the previous construction obtaining the so-called \emph{generalized Gabidulin codes}
\[\mathcal{G}_{k,s}=\langle x,x^{q^s},\ldots,x^{q^{s(k-1)}} \rangle_{\F_{q^n}},\]
with $\gcd(s,n)=1$ and $k\leq n-1$.
$\mathcal{G}_{k,s}$ is an $\F_{q}$-linear MRD-code with parameters $(n,n,q;n-k+1)$ and $L(\mathcal{G}_{k,s})=R(\mathcal{G}_{k,s})\simeq \F_{q^n}$, see \cite[Lemma 4.1 \& Theorem 4.5]{LN2016} and \cite[Theorem IV.4]{Morrison}.
Note that, as proved in \cite{Gabidulin,kshevetskiy_new_2005}, this family is closed by the Delsarte duality and by the adjoint operation, more precisely $\mathcal{G}_{k,s}^\perp$ is equivalent to $\mathcal{G}_{n-k,s}$ and $\mathcal{G}_{k,s}^\top$ is equivalent to itself.

More recently, Sheekey in \cite{Sheekey2016} proves that with $\gcd(s,n)=1$, the set
\[ \mathcal{H}_{k,s}(\eta,h)=\{a_0x+a_1x^{q^s}+\ldots+a_{k-1}x^{q^{s(k-1)}}+a_0^{q^h}\eta x^{q^{sk}} \colon a_i \in \F_{q^n}\}, \]
with $k\leq n-1$ and $\eta \in \F_{q^n}$ such that $\N_{q^n/q}(\eta)\neq (-1)^{nk}$, is an $\F_q$-linear MRD-code of dimension $nk$ with parameters $(n,n,q;n-k+1)$.
This code is called \emph{generalized twisted Gabidulin code}.
Lunardon, Trombetti and Zhou in \cite{LTZ}, generalizing the results of \cite{Sheekey2016}, determined the automorphism group of the generalized twisted Gabidulin codes and proved that, up to equivalence, the generalized Gabidulin codes and the twisted Gabidulin codes are both proper subsets of this class.
Clearly, for $\eta=0$ we have exactly the generalized Gabidulin code $\mathcal{G}_{k,s}$.
Also, the authors in \cite[Corollary 5.2]{LTZ} determined the left and right idealisers: if $\eta \neq 0$, then
\begin{equation}\label{leftrightidealH}
L(\mathcal{H}_{k,s}(\eta,h))\simeq\F_{q^{\gcd(n,h)}} \,\, \text{and} \,\, R(\mathcal{H}_{k,s}(\eta,h))\simeq\F_{q^{\gcd(n,sk-h)}}.
\end{equation}
The class of generalized twisted Gabidulin codes is closed by Delsarte duality and by the adjoint operation, more precisely $\mathcal{H}_{k,s}(\eta,h)^\perp$ is equivalent to $\mathcal{H}_{n-k,s}(-\eta,n-h)$ and $\mathcal{H}_{k,s}(\eta,h)^\top$ is equivalent to $\mathcal{H}_{k,s}(1/\eta,sk-h)$,
\cite[Theorem 6]{Sheekey2016} and \cite[Propositions 4.2 \& 4.3]{LTZ}.
We are interested to the case in which $h=0$, i.e.
\[  \mathcal{H}_{k,s}(\eta):=\mathcal{H}_{k,s}(\eta,0)=\langle x+\eta x^{q^{sk}}, x^{q^s},\ldots, x^{q^{s(k-1)}}\rangle_{\F_{q^n}}, \]
which is $\F_{q^n}$-linear (more precisely it is an $\F_{q}$-linear MRD-codes $\F_{q^n}$-linear on the left).
Apart from the two infinite families of $\F_{q}$-linear MRD-codes $\F_{q^n}$-linear on the left ($\cG_{k,s}$ and $\cH_{k,s}(\eta)$), there are a few other examples known for $n \in \{6,7,8\}$.

\begin{itemize}
  \item In \cite{CMPZ}, Csajb\'ok, Marino, Polverino and Zanella prove the following results
        \begin{itemize}
             \item for $q>4$ it is always possible to find $\delta \in
                   \F_{q^2}$ such that the set
                   \[ \C_1=\langle x,\delta x^{q}+x^{q^4} \rangle_{\F_{q^6}} \]
                   is an MRD-code with parameters $(6,6,q;5)$, \cite[Theorem 7.1]{CMPZ}. Its Delsarte dual code is equivalent to
                   \[ \D_1=\langle x^{q}, x^{q^{2}}, x^{q^{4}},x-\delta^{q} x^{q^3} \rangle_{\F_{q^6}}, \]
                    whose parameters are $(6,6,q;3)$;
            \item for $q$ odd and $\delta \in \F_{q^8}$ such that
                  $\delta^2=-1$ the set
                  \[ \C_2=\langle x,\delta x^{q}+x^{q^{5}} \rangle_{\F_{q^8}} \]
                  is an MRD-code with parameters $(8,8,q;7)$, \cite[Theorem 7.2]{CMPZ}. Its Delsarte dual code is is equivalent to
                  \[ \D_2=\langle x^{q},x^{q^2},x^{q^3},x^{q^5},x^{q^6},x-\delta x^{q^4} \rangle_{\F_{q^8}}, \]
                  whose parameters are $(8,8,q;3)$.
        \end{itemize}
  \item In \cite{CsMPZh}, Csajb\'ok, Marino, Polverino and Zhou prove the following results
        \begin{itemize}
          \item for $q$ odd and $\gcd(s,7)=1$ the set
                \[ \C_3=\langle x,x^{q^s}, x^{q^{3s}} \rangle_{\F_{q^7}} \]
                is an MRD-code with parameters $(7,7,q;5)$, \cite[Theorem 3.3]{CsMPZh}. Its Delsarte dual code is equivalent to
                \[ \D_3=\langle x,x^{q^{2s}},x^{q^{3s}},x^{q^{4s}} \rangle_{\F_{q^7}}, \]
                whose parameters are $(7,7,q;4)$;
          \item for $q \equiv 1 \pmod{3}$ and $\gcd(s,8)=1$ the set
                \[ \C_4=\langle x,x^{q^s}, x^{q^{3s}} \rangle_{\F_{q^8}} \]
                is an MRD-code with parameters $(8,8,q;6)$, \cite[Theorem 3.5]{CsMPZh}. Its Delsarte dual code is is equivalent to
                \[\D_4=\langle x,x^{q^{2s}},x^{q^{3s}},x^{q^{4s}},x^{q^{5s}} \rangle_{\F_{q^8}}\]
                whose parameters are $(8,8,q;4)$.
        \end{itemize}
  \item In \cite{CsMZ2018} Csajb\'ok, Marino and Zullo prove that for $q$ odd, $\delta^2+\delta=1$ and $q \equiv 0,\pm 1 \pmod{5}$
        \[\C_5=\langle x, x^{q}+x^{q^3}+\delta x^{q^5} \rangle_{\F_{q^6}}\]
        is an MRD-code with parameters $(6,6,q;5)$, \cite[Theorem 6.1]{CsMZ2018}. Its Delsarte dual code is is equivalent to
        \[ \D_5=\langle x^{q},x^{q^3},x-x^{q^2},x^{q^4}-\delta x \rangle_{\F_{q^6}} \]
        and it has parameters $(6,6,q;3)$.
\end{itemize}

\section{Characterization of generalized twisted Gabidulin codes}

In this section, we shall deal with $\F_q$-linear MRD-codes $\F_{q^n}$-linear on the left; hence from now on we will refer to them just as $\F_{q^n}$-linear MRD-codes.
Also, we may always assume that $\C\subseteq \tilde{\mathcal{L}}_{n,q}$ and that $L(\C)=\{\alpha x \colon \alpha \in \F_{q^n}\}$.
So, if $k$ is the dimension of $\C$ over $L(\C)$, we have that
\[ \C=\langle f_1(x),\ldots, f_k(x) \rangle_{\F_{q^n}}, \]
for some $f_1(x),\ldots, f_k(x) \in \tilde{\mathcal{L}}_{n,q}$.

Horlemann-Trautmann and Marshall in \cite{H-TM}
prove the following characterization of generalized Gabidulin codes.
In the original version, this result was formulated regarding RD-codes as subsets of $\F_{q^m}^n$; here we rewrite it in the context of linearized polynomials with $\C \subseteq \tilde{\mathcal{L}}_{n,q}$.

\begin{theorem}\label{gabidulind}\cite[Theorem 4.8]{H-TM}
  An $\F_{q^n}$-linear MRD-code $\C \subseteq \tilde{\mathcal{L}}_{n,q}$ having dimension $k$ is equivalent
  to a generalized Gabidulin code $\cG_{k,s}$ if and only if there is
  an integer $s<n$ with $\gcd(s,n)=1$ and
  $\dim_{\F_{q^n}} (\C\cap\C^{[s]})=k-1$, where $\C^{[s]}=\{f(x)^{[s]} \colon f(x)\in \C\}$,
  denoting by $[s]$ the power $q^s$.
\end{theorem}

Consider the subfamily of $\F_{q^n}$-linear generalized twisted Gabidulin codes
\[\cH_{k,s}(\eta)=\cH_{k,s}(\eta,0)=\langle x+\eta x^{[sk]}, x^{[s]},\ldots, x^{[s(k-1)]}\rangle_{\F_{q^n}},\]
with $\eta \neq 0$.
Note that if $\C\subseteq \tilde{\mathcal{L}}_{n,q}$ is an $\F_{q^n}$-linear MRD-code of dimension $k$ equivalent to a generalized twisted Gabidulin code $\cH_{k,s}(\eta)$ with $\eta\neq0$, then $\dim_{\F_{q^n}} (\C\cap\C^{[s]})=k-2$.
This condition, in general, is not
enough to characterize MRD-codes equivalent to $\cH_{k,s}(\eta)$.
In \cite{GiuZ}, we determine what further
conditions are necessary for a characterization in the spirit of Theorem \ref{gabidulind}, proving the following result.

\begin{theorem}\cite[Theorem 3.9]{GiuZ}
  Let $\C$ be a $k$-dimensional $\F_{q^n}$-linear MRD-code with $k>2$ contained
  in $\tilde{\mathcal{L}}_{n,q}$.
  Then, the code $\C$ is equivalent to a generalized twisted Gabidulin code if and only if there exists an integer $s$ such that $\gcd(s,n)=1$ and such that the following two conditions hold
\begin{enumerate}
  \item $\dim (\C \cap \C^{[s]})=k-2$ and $\dim(\C\cap \C^{[s]} \cap \C^{[{2s}]})=k-3$, i.e.
        there exist $p(x),q(x) \in \tilde{\mathcal{L}}_{n,q}$ such that
        \[ \C= \langle p(x)^{[s]}, p(x)^{[{2s}]}, \ldots, p(x)^{[{s(k-1)}]} \rangle_{\F_{q^n}} \oplus \langle q(x) \rangle_{\F_{q^n}}; \]
  \item $p(x)$ is invertible and there exists $\eta \in \F_{q^n}^*$ with $\N_{q^n/q}(\eta)\neq (-1)^{nk}$ such that $p(x)+\eta p(x)^{[{sk}]} \in \C$.
\end{enumerate}
\end{theorem}

\section{Distinguishers for RD-codes}

A \emph{distinguisher}\index{distinguisher} is an easy to compute function which allows to
identify an object in a family of (apparently) similar ones.
Existence of distinguishers is of particular interest for cryptographic
applications, as it makes possible to identify a candidate encryption
from a random text. Here, we shall present some new distinguishers for rank metric codes.

As seen in the previous section,
it has been shown in~\cite{H-TM} that an MRD-code $\C$ of parameters
$[n,k]$, i.e a $k$-dimensional MRD-code of $\tilde{\mathcal{L}}_{n,q}$, is equivalent
to a generalized Gabidulin code if, and only if, there exists a positive integer $s$ such that $\gcd(s,n)=1$ and $\dim(\C\cap\C^{[s]})=k-1$.
Following the approach of \cite{H-TM}, we define for any RD-code
$\C\subseteq \tilde{\mathcal{L}}_{n,q}$ the number
\begin{equation}\label{hind}
h(\C):=\max\{ \dim(\C\cap\C^{[j]})\colon j=1,\ldots,n-1; \gcd(j,n)=1 \}.
\end{equation}
Theorem~\ref{gabidulind} states that
an MRD-code $\C$ is equivalent to a generalized Gabidulin code
if and only if $h(\C)=k-1$.

We now define also the \emph{Gabidulin index}, $\mathrm{ind}(\C)$ of a $[n,k]$ RD-code
as the maximum dimension of a subcode $\mathcal{G}\leq\C$ contained in $\C$ with
$\mathcal{G}$ equivalent to a generalized Gabidulin code.
Note that the Gabidulin index is an invariant with respect to  strong equivalence since $\mathcal{G}_{k,s}^\top\simeq\mathcal{G}_{k,s}$.

In \cite{GiuZ}, we determine these indexes for each known $\F_{q^n}$-linear MRD-code.
Our result is contained in Table~\ref{tab1} and it is of interest since it can be used for the equivalence issue.
Note that in Table~\ref{tab1}, $\mathrm{ind}$, $h$ and $R$ denote, respectively, the Gabidulin index, the value defined in \eqref{hind} and the right idealiser.

  \begin{table}[htp]
    \[
      \begin{array}{ |c|c|c|c|c| }
        \hline
\mbox{Code } & \mathrm{ind} & h & R & [n,k] \\ \hline
\cG_{k,s} & k & k-1 & \F_{q^n} & [n,k] \\ \hline
\cH_{k,s}(\eta) & k-1 & k-2 & \F_{q^{\gcd(n,k)}} & [n,k] \\ \hline
\C_1 & 1 & 0 & \F_{q^3} & [6,2] \\ \hline
\C_2 & 1 & 0 & \F_{q^4} & [8,2] \\ \hline
\C_3 & 2 & 1 & \F_{q^7} & [7,3] \\ \hline
\C_4 & 2 & 1 & \F_{q^8} & [8,3] \\ \hline
\C_5 & 1 & 0 & \F_{q^2} & [6,2] \\ \hline
      \end{array}\qquad\qquad
      \begin{array}{ |c|c|c|c|c| }
        \hline
\mbox{Code } & \mathrm{ind} & h & R & [n,k] \\ \hline
 & & & & \\ \hline
 & & & & \\ \hline
\D_1 & 2 & 2 & \F_{q^3} & [6,4] \\ \hline
\D_2 & 3 & 4 & \F_{q^4} & [8,6] \\ \hline
\D_3 & 3 & 2 & \F_{q^7} & [7,4] \\ \hline
\D_4 & 4 & 3 & \F_{q^8} & [8,5] \\ \hline
\D_5 & 2 & 2 & \F_{q^2} & [6,4] \\ \hline
\end{array}
\]
\caption{Known linear MRD-codes and their distinguishers}
\label{tab1}
\end{table}

Clearly, the Gabidulin index may be defined also in the case in which the code considered is not $\F_{q^n}$-linear. For instance, we get the following.

\begin{theorem}
Let $n,s$ be positive integers such that $\gcd(s,n)=1$.
Let $k\leq n-1$ and $\eta \in \F_{q^n}^*$ such that $\N_{q^n/q}(\eta)\neq (-1)^{nk}$, then $\mathcal{H}_{k,s}(\eta,h)$ has Gabidulin index equals to $k-1$.
\end{theorem}

\smallskip
\noindent{\bf Proof}
\smallskip
As recalled in Section \ref{examples},
\[ \mathcal{H}_{k,s}(\eta,h)=\{a_0x+a_1x^{q^s}+\ldots+a_{k-1}x^{q^{s(k-1)}}+a_0^{q^h}\eta x^{q^{sk}} \colon a_i \in \F_{q^n}\}, \]
and since $\eta \neq 0$, it is not equivalent to any generalized Gabidulin code.
Hence, $\mathrm{ind}(\mathcal{H}_{k,s}(\eta,h))<k$. Also, it contains $\mathcal{G}'$, where
\[ \mathcal{G}'=\{ a_1x^{q^s}+\ldots+a_{k-1}x^{q^{s(k-1)}} \colon a_i \in \F_{q^n} \}\simeq \mathcal{G}_{k-1,s}. \]
Hence, $\mathrm{ind}(\mathcal{H}_{k,s}(\eta,h))=k-1$.

\medskip

We aim to continue in investigating such kind of results.

\vskip.2cm
\noindent
\begin{minipage}[t]{\textwidth}
Authors' addresses:
\vskip.2cm\noindent\nobreak
\centerline{
\begin{minipage}[t]{7cm}
Luca Giuzzi\\
D.I.C.A.T.A.M. {\small (Section of Mathematics)} \\
University of Brescia\\
Via Branze 43, I-25123, Brescia, Italy \\
luca.giuzzi@unibs.it
\end{minipage}
\begin{minipage}[t]{7.5cm}
Ferdinando Zullo\\
Department of Mathematics and Physics \\
University of Campania ``\emph{Luigi Vanvitelli}'' \\
Viale Lincoln 5, I-81100, Caserta, Italy \\
ferdinando.zullo@unicampania.it
\end{minipage}
}

\end{minipage}
\end{document}